\documentclass[twocolumn]{jpsj2} 
\title{The Effect of Oscillating Fermi Energy on the Line Shape of the Shubnikov-de Haas Oscillation in a Two-Dimensional Electron Gas}

\author{Akira \textsc{Endo}\thanks{E-mail address: akrendo@issp.u-tokyo.ac.jp} and Yasuhiro \textsc{Iye}}

\inst{Institute for Solid State Physics, University of Tokyo, 5-1-5 Kashiwanoha, Kashiwa, Chiba 277-8581}

\abst{The line shape of the Shubnikov-de Haas (SdH) oscillation has been analyzed in detail for a GaAs/AlGaAs two-dimensional electron gas. The line shape, or equivalently the behavior of the Fourier components, of the experimentally observed SdH oscillation is well reproduced by the sinusoidal density of states at the Fermi energy that oscillates with a magnetic field in a saw-tooth shape to keep the electron density constant. This suggests that the broadening of each Landau level by disorder is better described by a Gaussian than by a Lorentzian. }

\kword{Shubnikov-de Haas oscillation, two-dimensional electron gas, higher harmonics, Fermi energy, Gaussian, Lorentzian}

\begin{document}
\maketitle

\section{Introduction} 
The Shubnikov-de Haas (SdH) oscillation is a prevalently observed phenomenon in two-dimensional electron gases (2DEGs) subjected to a magnetic field perpendicular to the 2D plane. The oscillation is periodic in $1/B$ with the frequency proportional to the electron density $n_e$, thus serving as a standard tool to evaluate $n_e$ \cite{Stormer79}. The damping of the oscillation with decreasing magnetic field reflects the scattering of electrons out of the cyclotron orbit, and therefore is widely used as a measure of the single particle scattering time $\tau_\mathrm{Q}$ or the corresponding quantum mobility $\mu_\mathrm{Q} = e \tau_\mathrm{Q} / m^*$\cite{Coleridge91}, with $m^*$ the electron effective mass. 

Despite the ubiquity of the SdH oscillation in varieties of magnetoresistance experiments on a 2DEG, practically only the two aspects, the $1/B$ frequency and the damping, have been exploited so far to characterize the 2DEG used in the experiment. This requires   examining only the fundamental component of the oscillation, i.e., the sinusoidal oscillation with a due exponential damping factor. To the best of our knowledge, no report has been made to date on detailed investigation of the line shape, the behavior of higher harmonics, \cite{HigherHarm} of the SdH oscillation in a 2DEG \cite{NoteAsym}.

The SdH oscillation derives from the oscillating density of states (DOS) mainly through the modification of the scattering rate. The oscillation of the DOS, in turn, originates from the Landau quantization,
\begin{equation}
E_N = \hbar \omega_\mathrm{c} \left( N + \frac{1}{2} \right)
\label{EN}
\end{equation} 
with $\omega_\mathrm{c} = eB/m^*$, which turns the energy independent DOS at $B = 0$ into a set of delta function peaks at $E = E_N$ for an ideal 2DEG\@. The Landau level (LL) peaks, in reality, acquire width due to impurity scattering, which is usually modeled either by a Lorentzian $P_\mathrm{L} (E) = (\Gamma_\mathrm{L} /\pi )/(E^2  + \Gamma_\mathrm{L} ^2 )$, or by a Gaussian $P_\mathrm{G} (E) = (1/\sqrt {2\pi } \Gamma_\mathrm{G} )\exp ( - E^2 /2\Gamma_\mathrm{G} ^2 )$. The line shape of the resultant DOS, 
\begin{equation}
D(E) = \frac{2}{{2\pi l^2 }}\sum\limits_{N = 0}^\infty  {P_\alpha  } (E - E_N ),
\label{DOS}
\end{equation}
will certainly be reflected in the line shape of the SdH oscillation. In eq. (\ref{DOS}) we included the factor 2 to account for the spin degeneracy, $\alpha =$ L (G) for Lorentzian (Gaussian) peaks, and $l = \sqrt{\hbar/eB}$ represents the magnetic length. The analysis of the line shape of the SdH oscillation, therefore, allows us, in principle, to gain insight into the LL peaks that constitute the DOS\@.

Detailed knowledge of the DOS or the constituent LL peaks is indispensable to the quantitative understanding of a multitude of phenomena that originate from the Landau quantization. The two models of the LL broadening, the Lorentzian and the Gaussian, considerably differ at their tails. Therefore their difference can be crucial in the quantitative interpretation of the phenomena that take place at LL tails (e.g., localization in the quantum Hall states), as well as of those that occur in the low-magnetic field region where adjacent LLs substantially overlap.  A number of experimental techniques have been applied to the exploration of the DOS of a 2DEG, including the measurement of the specific heat \cite{Gornik85}, photoluminescence \cite{Berendschot87}, magnetocapacitance \cite{Ashoori92,Dial07}, and magnetization \cite{Eisenstein85,Potts96,Zhu03}. In comparison with these techniques, magnetoresistance measurement can be carried out with a simpler experimental setup thus generally with higher signal-to-noise (s/n) ratio, which is advantageous in investigating low magnetic-field range where the amplitude of the oscillation is expected to be small. A major drawback of the magnetoresistance in this respect is the possible intervention by effects other than that of the DOS; these include the weak localization effect, localization in the quantum Hall regime, formation of the edge states. These difficulties can mostly be circumvented by limiting ourselves to low magnetic fields where adjacent LLs have sufficient overlap to prevent the localized states in the bulk, hence also the edge states, from being generated. Weak localization, if any, is already suppressed at the magnetic field at which the SdH oscillation appears in the modern high-mobility GaAs/AlGaAs 2DEGs.

In the present paper, we make an analysis of the experimentally observed line shape of the low-field SdH oscillations. Comparison is made with the line shape of calculated DOS\@. Emphasis is on the importance to take into consideration the oscillation of the Fermi energy $E_\mathrm{F}$ that keeps the electron density constant. 

\section{Experimental Details}
We examined several GaAs/AlGaAs single-heterostructure 2DEG wafers with slightly varying sample parameters but all grown with the same molecular beam epitaxy (MBE) chamber. They all have a structure from the surface: 10 nm GaAs cap / 40 nm Si-doped ($\sim$10$^{24}$ m$^{-3}$) Al$_{0.33}$Ga$_{0.67}$As / undoped Al$_{0.33}$Ga$_{0.67}$As spacer having the thickness $d_\mathrm{s} =$ 40 or 60 nm / GaAs with 2DEG channel residing at the interface with the spacer layer. The electron density $n_e$ and the mobility $\mu$ are 1.5$-$2.0$\times$10$^{15}$ m$^{-2}$ and 50$-$70 m$^2$/Vs, respectively, in the dark, which increase to 2.2$-$3.0 $\times$ 10$^{15}$ m$^{-2}$ and 70$-$130 m$^2$/Vs after illumination by an infrared light emitting diode. The samples are fabricated into Hall bars for magnetotransport measurements. All the measured SdH traces taken from different samples or with different illumination conditions showed essentially the same features, with minor variations to be mentioned when necessary. In what follows, we present a typical example taken at the conditions $n_e =$ 2.9$\times$10$^{15}$ m$^{-2}$ and $\mu =$ 74 m$^2$/Vs obtained after illumination.

The measurement was carried out in a dilution fridge at the base temperature ($\sim$ 15 mK). We employed standard low-frequency (13 Hz) ac lock-in technique for resistance measurement, with an excitation current of $I_\mathrm{rms} =$ 10 nA; the only effect found by reducing $I_\mathrm{rms}$ down to 0.5 nA was to worsen the s/n ratio, attesting to the absence of the current heating. In order to capture the finest details of the oscillation, we adopted a slow sweep rate (0.01 T/min) of the magnetic field and a rapid data acquisition rate ($\sim$ 4 data points/s), which amounts to roughly 1 data point per 4$\times$10$^{-5}$ T\@. 
The slow sweep rate is also beneficial in avoiding undesirable hysteresis of the superconducting magnet. The applied magnetic fields were further calibrated by the simultaneously measured Hall resistivity.

\section{Experimental Results}

\begin{figure}[tb]
\includegraphics[bbllx=20,bblly=20,bburx=550,bbury=370,width=8.5cm]{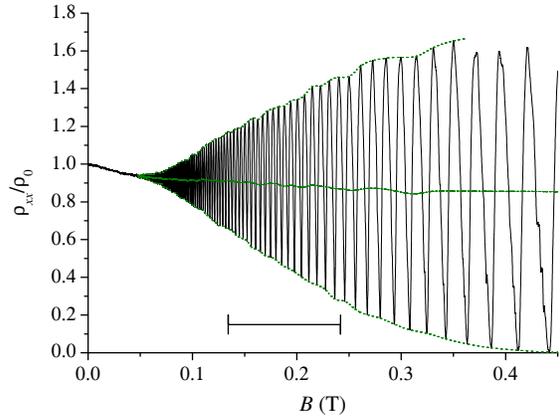}
\caption{(Color online) A low-field magnetoresistance trace of a 2DEG measured at $\sim$ 15 mK\@. Upper/lower envelope curves and their average are shown by dotted lines and a dot-dashed line, respectively. The horizontal bar indicates the range of the magnetic field plotted in Fig.\ \ref{expSdH}.}
\label{rawF}
\end{figure}

\begin{figure}[tb]
\includegraphics[bbllx=20,bblly=20,bburx=550,bbury=360,width=8.5cm]{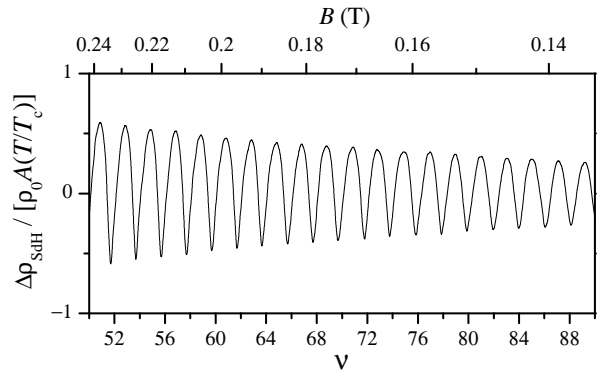}
\caption{The SdH oscillation extracted from Fig.\ \ref{rawF}, plotted against the filling factor. The corresponding magnetic field is indicated by the top axis.}
\label{expSdH}
\end{figure}

\begin{figure}[tb]
\includegraphics[bbllx=20,bblly=20,bburx=550,bbury=360,width=8.5cm]{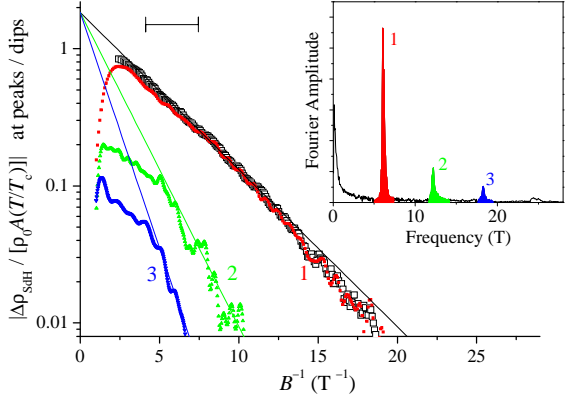}
\caption{(Color online) Inset: the Fourier spectrum obtained from the $\Delta \rho_\mathrm{SdH} / [\rho_0 A(T/T_\mathrm{c})]$ vs. $1/B$ curve. Main panel: the amplitudes of the fundamental (red squares), the second (green upward triangles), and the third (blue downward triangles) harmonic contents of the SdH oscillation [see Fig.\ \ref{BPF}(a)] plotted against $1/B$. Amplitudes of the total SdH oscillation (Fig.\ \ref{expSdH}) are also plotted by open squares. The line with the least steep slope represents the fit of $C \exp(-\pi / \mu_\mathrm{Q}B)$ to the total SdH amplitude (after omitting the higher magnetic-field range), which defines $\mu_\mathrm{Q}$=11.9 m$^2$/Vs. The two steeper lines show $C \exp(-2 \pi / \mu_\mathrm{Q}B)$ and $C \exp(-3 \pi / \mu_\mathrm{Q}B)$ with the same values of $C$ and $\mu_\mathrm{Q}$ as the first line. The horizontal bar indicates the magnetic-field range plotted in Figs.\ \ref{expSdH} and \ref{BPF}.}
\label{ampFFT}
\end{figure}

\begin{figure}[tb]
\includegraphics[bbllx=20,bblly=20,bburx=550,bbury=800,width=8.5cm]{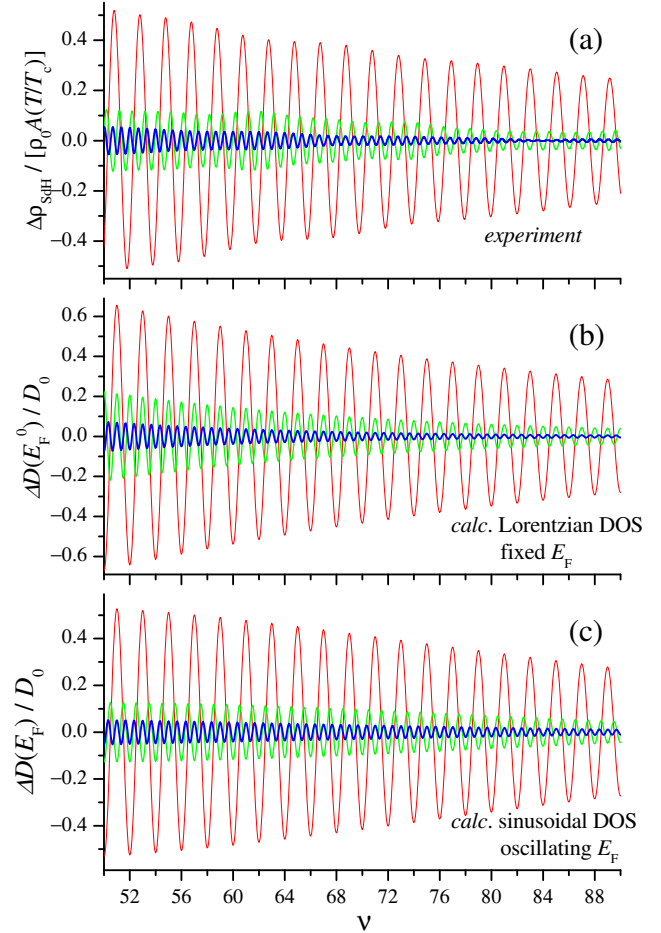}
\caption{(Color online) (a) The fundamental (red), the second (green), and the third (blue) harmonic content of the SdH oscillation (plotted with progressively thicker line) obtained by a numerical Fourier band pass filter performed on $\Delta \rho_\mathrm{SdH} / [\rho_0 A(T/T_\mathrm{c})]$ shown in Fig.\ \ref{expSdH}. The frequency windows allowed by the filter are indicated by the shade in the inset of Fig.\ \ref{ampFFT}. (b)(c) Similar to (a) with the band pass filter performed  on the Lorentzian DOS at fixed $E_\mathrm{F}$ [Fig.\ \ref{calcLorentz}(a)] (b), or on the sinusoidal DOS at oscillating $E_F$ [Fig.\ \ref{calcsin}(b)] (c). Note the difference in the phase of the second harmonic component between (b) and (a), (c). See text for detail.}
\label{BPF}
\end{figure}

Figure \ref{rawF} shows a typical low-field magnetoresistance trace, with $x$ the direction of the current and $\rho_0$ the resistivity $\rho_{xx}$ at $B =$ 0. The amplitude of the SdH oscillation monotonically increase with increasing $B$ up until $\sim$0.3 T\@. The deviation from this trend at higher magnetic fields is attributable to the onset of the spin splitting. In the present paper, we examine only the low magnetic-field region where spin degeneracy remains unresolved.  Note that there the bare Zeeman splitting is only a few percent of the disorder broadening $\Gamma_\alpha$ of the LLs owing to the smallness of the $g$ factor in GaAs \cite{gfactor}, and is therefore completely negligible. The spin splitting that evolves into the odd-integer quantum Hall states at higher magnetic fields is predominantly due to the exchange enhancement of the $g$ factor that takes place only above a certain magnetic field; the onset is considered to be a second-order phase transition \cite{Fogler95} and consequently the exchange enhancement is irrelevant below the transition field. In Fig.\ \ref{rawF}, we also plot upper and lower envelope curves (dotted lines) and their average (a dot-dashed line). The envelope curves are obtained as spline curves connecting the extrema. The oscillatory part of the magnetoresistance $\Delta \rho_\mathrm{SdH} / \rho_0$ is extracted from $\rho_{xx} / \rho_0$ by subtracting the average curve. The resultant SdH oscillation $\Delta \rho_\mathrm{SdH} / [\rho_0 A(T/T_\mathrm{c})]$ is plotted in Fig.\ \ref{expSdH} against the LL filling factor $\nu = n_e h /eB$ for the magnetic-field range depicted by a horizontal bar in Fig.\ \ref{rawF}. The factor $A(T/T_\mathrm{c})$ with $k_\mathrm{B} T_\mathrm{c} = (1/2 \pi^2) \hbar \omega_\mathrm{c}$ and $A(x) = x / \sinh(x)$ corrects for thermal damping; the decrement of the factor from unity is less than 0.01 for the relevant magnetic-field range, indicating that thermal damping is negligibly small for our low temperature.

The SdH trace in Fig.\ \ref{expSdH} takes minima and maxima at even and odd filling factors, respectively, as is expected for a spin-degenerate 2DEG\@. A notable feature to be highlighted in the present paper is the asymmetry between the maxima and the minima. While the peaks at the maxima exhibit rather dull rounded inverted U-shape, the dips at minima take on sharp V-shape. This trait in the line shape is a quite generic feature observed in all the 2DEGs we have investigated and also can be seen in the papers published by other authors (see, e.g., Fig.\ 1 in Ref.\ \citen{Coleridge94}).

In order to quantify the line shape, we carried out Fourier transform to the trace of $\Delta \rho_\mathrm{SdH} / [\rho_0 A(T/T_\mathrm{c})]$ vs. $1/B$ (note that $1/B \propto \nu$ so that the trace to be Fourier transformed is basically the same as the trace shown in Fig.\ \ref{expSdH}). The Fourier spectrum shown in the inset of Fig.\ \ref{ampFFT} exhibits peaks deriving from the fundamental periodicity and up to the third (and a small trace of the fourth) harmonics. The resistivity component corresponding to each Fourier peak (up to the third harmonic) is obtained by performing a Fourier band pass filter to the $\Delta \rho_\mathrm{SdH} / [\rho_0 A(T/T_\mathrm{c})]$ vs. $1/B$ curve using the shaded region in the Fourier spectrum as a window, and plotted in Fig.\ \ref{BPF}(a) against $\nu$. Addition of the three traces in Fig.\ \ref{BPF}(a) practically reproduces the trace in Fig.\ \ref{expSdH}.

The amplitudes (absolute values of maxima/minima) of each oscillatory components in Fig.\ \ref{BPF}(a) are plotted in the main panel of Fig.\ \ref{ampFFT} (solid symbols) in the semi-logarithmic scale, along with the amplitudes obtained from Fig.\ \ref{expSdH} (open squares). It is clear from the figure that the SdH oscillation is dominated by the fundamental component. This justifies the conventional treatment in which the oscillation is approximated by a single sinusoidal curve with an exponential damping factor
\begin{equation}
\frac{\Delta \rho_\mathrm{SdH}}{\rho_0 A(T/T_\mathrm{c})} \simeq -C \cos \left( \pi \nu \right) \exp \left( -\frac{\pi}{\mu_\mathrm{Q} B} \right).
\label{Drho1}
\end{equation}
The quantum mobility $\mu_\mathrm{Q}$ is deduced from the damping of the amplitude. We obtain $\mu_\mathrm{Q}=$ 11.9 m$^2$/Vs by the fit of $C \exp (-\pi / \mu_\mathrm{Q} B)$ to the total SdH amplitudes, which fits to the amplitudes of the fundamental component as well, as seen in Fig.\ \ref{ampFFT}. The higher magnetic-field regime where spin-splitting commences is omitted from the fitting. Interestingly, the amplitudes of the second and the third harmonics roughly fall on the lines $C \exp (-2 \pi / \mu_\mathrm{Q} B)$ and $C \exp (-3 \pi / \mu_\mathrm{Q} B)$, respectively, with the identical values of $C$ and $\mu_\mathrm{Q}$ obtained by the above fitting. This property is shared by the SdH traces taken at other illumination conditions (i.e., different $\mu$ and $n_e$) and with other samples \cite{NoteCmu}. As will be discussed later, this behavior of the \textit{amplitude} is what we expect for a Lorentzian line shape, although the Lorentzian fails to reproduce the correct \textit{phase} of the second harmonic. In the subsequent section, we compare our experimental line shape with two types of calculated DOS\@.

\section{Comparison with Calculated DOS}
\subsection{Oscillation of the Fermi energy}
For low magnetic fields, it is convenient to rewrite, with the aid of the Poisson sum formula \cite{Ishihara86,Raikh93}, the DOS eq.\ (\ref{DOS}) in the Fourier series,
\begin{eqnarray}
\displaystyle{ \lefteqn{D(E) =}} \nonumber \\
 & \displaystyle{ D_0 \Biggl\{ 1 + 2\sum\limits_{k = 1}^\infty  \cos \left[ 2 \pi k \left( \frac{E}{\hbar \omega_\mathrm{c}} - \frac{1}{2}\right) \right] \Biggr. } \nonumber \\
 & \hspace{35mm} \displaystyle{ \Biggl. \times \exp \left[ - 2\left( \frac{\pi \Gamma_\alpha}{\hbar \omega_\mathrm{c} } k \right)^p \right] \Biggr\} ,}
\label{DOSP}
\end{eqnarray}
with $D_0 = m^* / \pi \hbar^2 $ the constant DOS at $B=0$, and $p =$ 1 for Lorentzian and 2 for Gaussian LL peaks. Here we resorted to an approximation $\sum_{N=0}^\infty$ $\rightarrow$ $\sum_{N=-\infty}^\infty$; the approximation causes virtually no change in the DOS at $E \gg \Gamma_\alpha$, i.e., at $E$ beyond the reach of the tail from the peaks located at $N <$ 0. Since terms with larger $k$ decay more rapidly owing to the exponential damping factor, it is usually enough to take only a few terms into account in the summation of eq.\ (\ref{DOSP}) for low magnetic fields. 

Most of the LL peaks experimentally measured so far have been explained either by a Lorentzian with the $B$-independent width \cite{Ashoori92,Potts96} or by a Gaussian with the width proportional to $\sqrt{B}$ \cite{Eisenstein85}. With 
\begin{equation}
\Gamma_\mathrm{L}  = \frac{e\hbar}{2m^*} \frac{1}{\mu _\mathrm{Q}}
\label{GammaL}
\end{equation}
and
\begin{equation}
\Gamma_\mathrm{G}  = \frac{e\hbar}{2m^*} \sqrt {\frac{2 B}{\pi \mu _\mathrm{Q}}},
\label{GammaG}
\end{equation}
the exponential factor in eq.\ (\ref{DOSP}) reduces to $\exp (-  k^p \pi / \mu_\mathrm{Q} B)$, and by substituting $k =$ 1 coincides with the exponential factor in eq.\ (\ref{Drho1}) for both Lorentzian and Gaussian.

Using the $D(E)$ in eq.\ (\ref{DOSP}), the electron density at low temperatures is written as
\begin{equation}
n_e  = \int_0^\infty  {D(E)f(E)dE} \simeq \int_0^{E_\mathrm{F}}  {D(E)dE} = N(E_\mathrm{F}) 
\label{density}
\end{equation}
with $f(E) = \{ 1+\exp[(E-E_\mathrm{F})/k_\mathrm{B}T] \}^{-1}$ the Fermi-Dirac distribution function. We defined the cumulative number of states (i.e., integrated DOS) below an energy $E$ as
\begin{eqnarray}
\lefteqn{\displaystyle{ N(E) = \int_0^E {D(E ^\prime)dE ^\prime}}} & \nonumber \\
 & \displaystyle{ = D_0 \Biggl\{ E + 2\sum\limits_{k = 1}^\infty  \frac{\hbar \omega _\mathrm{c} }{2\pi \mathrm{k} } \sin \left[ 2\pi k\left( \frac{E}{\hbar \omega _\mathrm{c}} - \frac{1}{2} \right) \right] \Biggr. }   \nonumber \\
 & \hspace{35mm} \displaystyle{ \Biggl. \times \exp \left[ - 2\left( \frac{\pi \Gamma }{\hbar \omega _\mathrm{c} } k \right)^p  \right]  \Biggr\} }.
\label{NE}
\end{eqnarray}
The electron density $n_e$ of a 2DEG is expected not to vary with magnetic field at low temperatures. To keep $n_e$ constant, eq.\ (\ref{density}) requires $E_\mathrm{F}$ to oscillate with $B$ around the value at $B =$ 0, $E_\mathrm{F}^0  = n_e /D_0 $. The oscillation of $E_F$ with $B$, as well as the DOS at the oscillating $E_\mathrm{F}$, is evaluated for the two types of LL broadening in the following subsections.

\subsection{Lorentzian density of states}
\label{subsecLorentzian}
\begin{figure}[tb]
\includegraphics[bbllx=20,bblly=40,bburx=550,bbury=760,width=8.5cm]{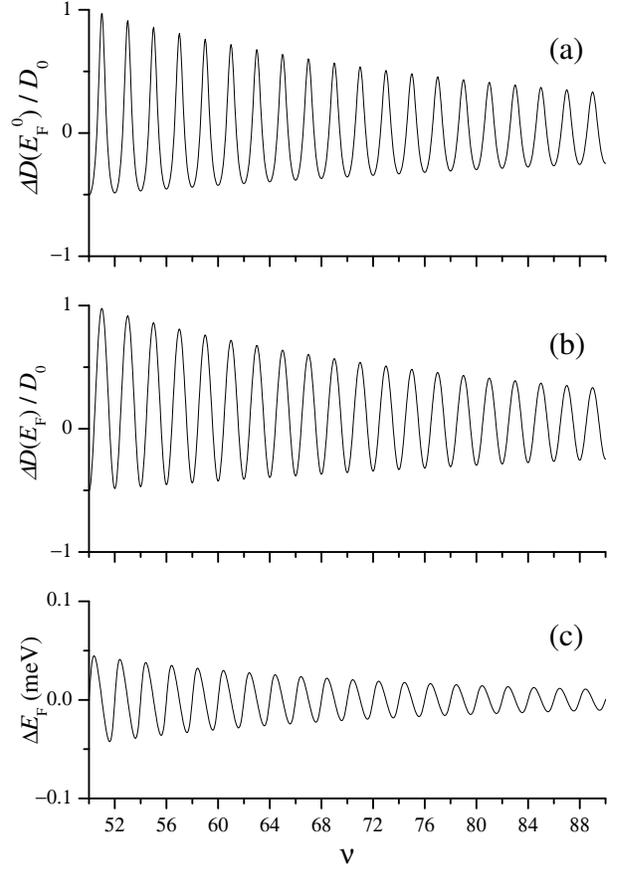}
\caption{The upper two panels show the oscillatory part of the Lorentzian DOS, with the width $\Gamma_\mathrm{L}$ determined from the experimentally obtained $\mu_\mathrm{Q} =$11.9 m$^2$/Vs. (a) assumes a fixed $E_\mathrm{F}^0 =$ 10.4 meV (corresponding to the experimental $n_e=$ 2.92$\times$10$^{15}$ m$^{-2}$), while (b) takes account of the oscillating $E_\mathrm{F}$ calculated by the procedure described in the text. (c) shows asymmetric (sawtooth-like) oscillation of the Fermi energy, repeating alternatingly steeper rises and gradual declines with $\nu$.}
\label{calcLorentz}
\end{figure}

The oscillatory part $\Delta D / D_0 = (D-D_0) / D_0$ of the Lorentzian DOS at fixed Fermi energy $E_\mathrm{F}^0$ is plotted in Fig.\ \ref{calcLorentz}(a). The line shape with sharp peaks and rounded dips is at obvious variance with the experimental line shape of the SdH oscillation. Equation (\ref{DOSP}) along with eq.\ (\ref{GammaL}) shows that the amplitude of the $k$-th harmonic is given by $2 \exp(-k \pi/\mu_\mathrm{Q} B)$. Therefore the decay with $1/B$ of the amplitude of the harmonics for the experimental SdH oscillation presented in the previous section (see Fig.\ \ref{ampFFT}) is in accord with that of the Lorentzian. However, the oscillation of each harmonic content shown in Fig.\ \ref{BPF}(b) obtained by the Fourier band pass filter reveals that the phase of the second harmonic of the Lorentzian DOS is inverted from that of the experimental trace. Therefore, the experimental SdH oscillation is more like a Lorentzian laid upside-down, $\Delta D \rightarrow - \Delta D$ after a shift $E_\mathrm{F}^0 \rightarrow E_\mathrm{F}^0 + \hbar \omega_c/2$ [see eq.\ (\ref{DOSP})]; it looks as if the gaps between LL peaks are comprised of inverted Lorentzians.

As discussed in the previous subsection, it is necessary to take the oscillation of the Fermi energy into account. For the Lorentzian DOS, $\Delta E_\mathrm{F}$ can be calculated analytically. Equation (\ref{DOSP}) with $p =$ 1 can further be rewritten as
\begin{equation}
D(\varepsilon) = D_0 \frac{\sinh (2\pi \gamma )}{\cosh (2\pi \gamma ) - \cos (2\pi \varepsilon )}.
\label{DOSL}
\end{equation}
Here we introduced dimensionless parameters, $\gamma = \Gamma_\mathrm{L}/\hbar \omega_\mathrm{c} = \pi (\Gamma_\mathrm{G}/\hbar \omega_\mathrm{c})^2 = (2 \mu_\mathrm{Q} B)^{-1} $ and $\varepsilon = E/\hbar \omega_\mathrm{c} - 1/2$. In what follows, we will also use dimensionless Fermi energies $\varepsilon_\mathrm{F}$ and $\varepsilon_\mathrm{F}^0$ similarly defined as $\varepsilon$, and also $\delta = \Delta E_\mathrm{F}/\hbar \omega_\mathrm{c}$ for the deviation $\Delta E_\mathrm{F} = E_\mathrm{F}- E_\mathrm{F}^0$ of the Fermi energy from the zero-field value. Note that $\varepsilon_\mathrm{F}^0 = \nu/2 - 1/2$. Accordingly, the cumulative number of states reads
\begin{eqnarray}
\lefteqn{N(\varepsilon) = }\nonumber \\
 & \displaystyle{ 2\frac{{eB}}{h}\left\{ {\frac{1}{\pi }\arctan \left[ {\coth (\pi \gamma )\tan (\pi \varepsilon )} \right] + \frac{1}{2} + \mathrm{Int}\left[ {\varepsilon  + \frac{1}{2}} \right]} \right\} }, \nonumber \\
\label{NEL}
\end{eqnarray}
where Int[$x$] signifies the integer part of $x$. From eqs.\ (\ref{density}) and (\ref{NEL}), one obtains
\begin{eqnarray}
\tan (\pi \varepsilon_\mathrm{F} ^0 ) & = &\coth (\pi \gamma )\tan (\pi \varepsilon_\mathrm{F} ) \nonumber \\
 & = & \coth (\pi \gamma )\tan \left[ {\pi (\varepsilon_\mathrm{F} ^0  + \delta )} \right],
\label{tantan}
\end{eqnarray}
leading to the oscillatory part of the Fermi energy
\begin{equation}
\delta  =  \frac{1}{\pi }\arctan \left\{ {\frac{{\left[ {\tanh (\pi \gamma ) - 1} \right]\tan (\pi \varepsilon_\mathrm{F} ^0 )}}{{1 + \tanh (\pi \gamma )\tan ^2 (\pi \varepsilon _\mathrm{F} ^0 )}}} \right\}.
\label{deltaL}
\end{equation}
The resulting $\Delta E_\mathrm{F}$ is plotted in Fig.\ \ref{calcLorentz}(c). From eq.\ (\ref{tantan}) one obtains
\begin{eqnarray}
\cos(2 \pi \varepsilon_\mathrm{F}) & = & \displaystyle{ \frac{1-\tan ^2 (\pi \varepsilon_\mathrm{F})}{1+\tan ^2 (\pi \varepsilon_\mathrm{F})} } \nonumber \\
& =& \displaystyle{ \frac{1-\tanh ^2 (\pi \gamma) \tan ^2 (\pi \varepsilon_\mathrm{F}^0)}{1+\tanh ^2 (\pi \gamma) \tan ^2 (\pi \varepsilon_\mathrm{F}^0)} }.
\label{costan}
\end{eqnarray}
The DOS at oscillating $E_F$ is obtained by replacing eq.\ (\ref{costan}) into eq.\ (\ref{DOSL})
\begin{equation}
D(\varepsilon_\mathrm{F}) = D_0 \left[ \coth(2 \pi \gamma)+\frac{1}{\sinh(2 \pi \gamma)} \cos(2 \pi \varepsilon_\mathrm{F}^0) \right],
\label{DEFLor}
\end{equation}
whose oscillatory part is plotted in Fig.\ \ref{calcLorentz}(b). Interestingly, higher harmonics are gone and the oscillation includes only the fundamental component. To conclude this subsection, the Lorentzian DOS does not reproduce the line shape of the experimentally observed SdH oscillation, even with the oscillatory $E_\mathrm{F}$.

\subsection{Sinusoidal density of states}

\begin{figure}[tb]
\includegraphics[bbllx=20,bblly=40,bburx=550,bbury=760,width=8.5cm]{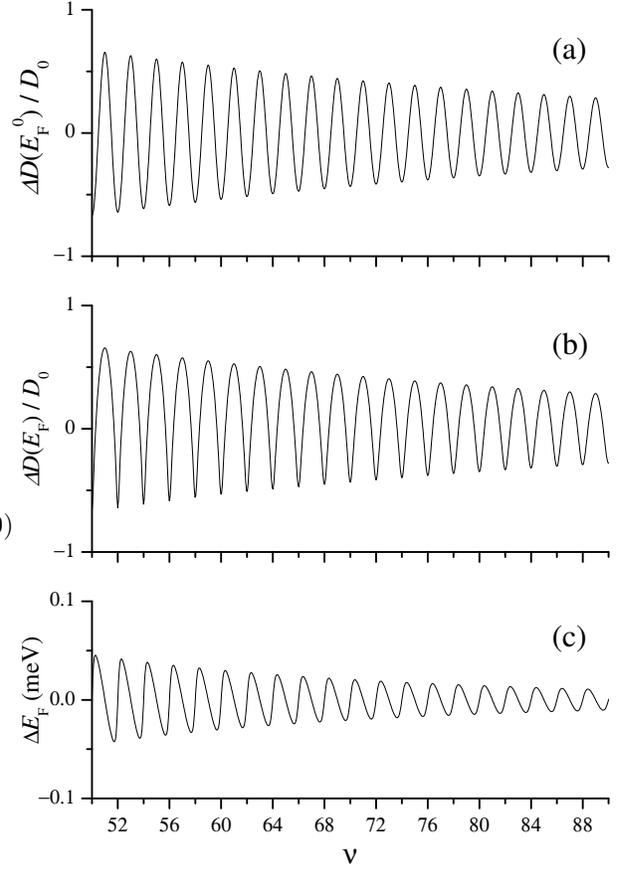}
\caption{Similar to Fig.\ \ref{calcLorentz} for the sinusoidal DOS, which serves as a good approximation for the Gaussian DOS as explained in the text.}
\label{calcsin}
\end{figure}

For the Gaussian DOS, it is rather difficult to perform similar calculations as were done in the previous subsection for the Lorentzian DOS\@. Instead, we study the sinusoidal DOS in this subsection. Since the terms with $k \geq$ 2 in eq.\ (\ref{DOSP}) decay rapidly for the Gaussian DOS because of the factor $k^2$ in the index of the exponential damping factor, the sinusoidal DOS that keeps only the term $k =$ 1 in the summation of eq.\ (\ref{DOSP}) constitutes a good approximation for the Gaussian DOS\@. Actually, the amplitude of the term $k =$ 2 accounts for only 0.1$-$3.8 \% of that of the term $k =$ 1 for the current example in the magnetic-field range of the present interest. For a sinusoidal DOS
\begin{equation}
D(\varepsilon) = D_0 \left[ 1 + 2 \cos(2 \pi \varepsilon) \exp(- 2 \pi \gamma) \right]
\label{DOSSin}
\end{equation}
we obtain
\begin{equation}
N(\varepsilon) = 2 \frac{eB}{h} \left[\varepsilon + \frac{1}{2}+\frac{1}{\pi} \sin(2 \pi \varepsilon) \exp(- 2 \pi \gamma) \right].
\label{NESin}
\end{equation}
Equations (\ref{density}) and (\ref{NESin}) result in
\begin{equation}
0 = \delta  + \frac{1}{\pi }\sin \left[ {2\pi \left( {\varepsilon _{\rm{F}} ^0  + \delta } \right)} \right] \exp (- 2 \pi \gamma). 
\label{deltaeq}
\end{equation}
Equation (\ref{deltaeq}) may be solved numerically. Here, instead, we deduce an approximate solution valid up to $O(\delta^2)$, noting the smallness of $\delta$ in the magnetic-field range of the current interest. We get
\begin{equation}
\delta \simeq \frac{1 + 2 \lambda c^0 - \sqrt{(1+2 \lambda c^0)^2 + 2 (2 \lambda s^0)^2}}{4 \pi \lambda s^0},
\label{deltaSin}
\end{equation}
where we have introduced notations $\lambda = \exp (- 2 \pi \gamma)$, $c^0 = \cos (2 \pi \varepsilon_\mathrm{F}^0)$, and $s^0 = \sin (2 \pi \varepsilon_\mathrm{F}^0)$ for brevity. The calculated $\Delta E_\mathrm{F}$ is plotted in Fig.\ \ref{calcsin}(c) along with the oscillatory part of the DOS either at fixed $E_\mathrm{F}^0$ or at oscillating $E_\mathrm{F}$ in Figs.\ \ref{calcsin}(a) and \ref{calcsin}(b), respectively \cite{Notenumcalc}. Figure \ref{calcsin}(b) reproduces the line shape of the experimental SdH oscillation in Fig.\ \ref{expSdH} quite well. Both the amplitude and the phase of the harmonics are basically reproduced, as can be discerned by comparing Figs.\ \ref{BPF}(a) and \ref{BPF}(c), the latter showing the harmonic contents of Fig.\ \ref{calcsin}(b) obtained by the Fourier band pass filter.

To be more quantitative, we deduce approximate formula for $\Delta D / D_0$ at oscillating $E_\mathrm{F}$ up to $O(\lambda^3)$. Note that the sinusoidal DOS starts to deviate from the Gaussian DOS only at $O(\lambda^4)$. Substituting 
\begin{equation}
2\pi \delta  \simeq  - 2\sin (2\pi \varepsilon _\mathrm{F}^0) \lambda + 2\sin (4\pi \varepsilon _\mathrm{F}^0)\lambda^2  + O(\lambda^3)
\nonumber
\end{equation}
into
\begin{equation}
\frac{\Delta D (\varepsilon_\mathrm{F})}{D_0} = 2\cos (2\pi \varepsilon _\mathrm{F}) \lambda = 2\cos \left[ 2\pi (\varepsilon _\mathrm{F}^0+\delta) \right] \lambda,
\label{DOSsinosc}
\end{equation}
one obtains
\begin{eqnarray}
\frac{\Delta D (\varepsilon_\mathrm{F})}{D_0} & = & 2\cos (2\pi \varepsilon _\mathrm{F}^0) \lambda + 2 \left[ 1 - \cos (4\pi \varepsilon _\mathrm{F} ^0) \right] \lambda^2 \nonumber \\
 & & + 3 \left[  - \cos (2\pi \varepsilon _\mathrm{F}^0) + \cos (6\pi \varepsilon _\mathrm{F}^0) \right] \lambda^3 + O(\lambda^4 ). \nonumber \\
\label{DOAsin}
\end{eqnarray}
Equation (\ref{DOAsin}) gives the right amplitude and phase of the harmonics as inferred from Figs.\ \ref{ampFFT} and \ref{BPF}(a), except for the 1.5 times larger amplitude of the third harmonic. The behavior of high harmonics, however, is rather subtle. In fact, the use of eq.\ (\ref{deltaL}) instead of eq.\ (\ref{deltaSin}), a fairly good approximation considering the minuteness of the difference between Figs.\ \ref{calcLorentz}(c) and \ref{calcsin}(c), leads to the expected amplitude and phase up to the third harmonic (see Appendix).

\section{Discussion}
Different experimental approaches employed to explore the LL line shape \cite{Gornik85,Berendschot87,Ashoori92,Dial07,Eisenstein85,Potts96,Zhu03} differ in the preference of Lorentzian or Gaussian. Recent experiments \cite{Ashoori92,Dial07,Potts96,Zhu03}, however, seem to converge on the Lorentzian broadening with $\Gamma_\mathrm{L}$ independent of $B$ for middle to high magnetic fields, $B > \sim$1 T\@. For lower magnetic fields, the result of recent high sensitivity magnetization measurement carried out down to a magnetic field as low as $\sim$0.5 T can be fitted by the two types of broadening equally well, hence cannot distinguish between the two \cite{Zhu03}. Therefore there has been no general agreement on the LL line shape at low magnetic fields. The present study suggests that the Gaussian is the better candidate for still lower magnetic fields. 

Difficulty in magnetization experiments at low magnetic fields appears to be arising at least partly from the limit in the sensitivity that hampers the acquisition of the data with sufficient s/n ratio. In magnetoresistance measurements, small amplitude oscillations can be detected with satisfactory s/n ratio as we have shown in the present study. On the other hand, the interpretation of magnetoresistance data becomes complicated at higher magnetic fields, affected by localization or edge states in the quantum Hall regime. Therefore the two experimental techniques are more or less complementary. 

Theoretically, a pioneering work by Ando and Uemura \cite{AndoLB74} suggested semi-elliptic LLs with the width proportional to $\sqrt{B}$ by self-consistent Born approximation (SCBA). The approximation is insufficient at the tail of LLs where multiple scattering plays important role, and the semi-elliptic LLs are not applicable at low magnetic fields where adjacent LLs overlap. Alternative approaches showed that the LL broadening is described by a Gaussian \cite{Gerhardts76,Wegner83}. In these early studies, a short-range random potential is assumed. A later work \cite{Raikh93} showed that a Gaussian line shape with the width $\Gamma_\mathrm{G} \propto \sqrt{B}$ holds also for a long-range random potential more appropriate for GaAs/AlGaAs 2DEGs. Our result is, therefore, in accordance with the theoretical prediction.

In the comparison between the experimental SdH oscillation and the calculated DOS, we have tacitly assumed the proportionality $\Delta \rho_\mathrm{SdH} / [\rho_0 A] \propto \Delta D /D_0$. It has been claimed, with ample experimental evidence, that $\rho_{xx}$ is proportional to the \textit{square} of the DOS \cite{Coleridge89,Coleridge97}. This leads to the same proportionality as long as the oscillatory part $\Delta D$ of the DOS is much smaller than the constant background $D_0$ so that an approximation $(D/D_0)^2 \simeq 1 + 2 \Delta D / D_0$ is allowed. However, $\Delta D / D_0$ is not necessarily small enough even in the low magnetic-field range examined in the present study, as can be seen in Figs.\ \ref{calcLorentz} and \ref{calcsin}. Therefore the relation $\rho_{xx} \propto D^2$ suggests that the comparison should be made between the experimental $\rho_\mathrm{SdH} / [\rho_0 A]$ and the oscillatory part of $D^2$. We have actually made such comparison, only to find much worse agreement with the experimental SdH oscillation for any combinations of the type of DOS and $E_\mathrm{F}$ (either fixed or oscillating). The relation $\rho_{xx} \propto D^2$ derives from the theory for short-range random potential \cite{Ishihara86}. The extension of the theory to the system with long-range random potential is not straightforward, involving subtleties in the treatment of relevant scattering times \cite{Coleridge89}. Our result rather suggests the relation $\Delta \rho_\mathrm{SdH} / [\rho_0 A] \propto \Delta D /D_0$ remains valid regardless of the magnitude of $\Delta D /D_0$. However, the possibility that this trait is specific to the samples we examined cannot be completely ruled out, since they are all grown in the same MBE chamber and therefore can possibly contain an unidentified common source of scattering unintentionally introduced during the growth.

\section{Conclusions}
We have made detailed Fourier analysis of the line shape of the SdH oscillation at low magnetic-field regime where the oscillation predominantly reflects the oscillation in the spin-degenerate DOS\@. The line shape can formally be described by inverted-Lorentzian gaps between adjacent LLs, and can be reproduced by the sinusoidal DOS at the $E_\mathrm{F}$ oscillating in a saw-tooth shape. The sinusoidal DOS is consistent with the Gaussian broadening of LLs with the width $\propto \sqrt{B}$, in agreement with the broadening theoretically calculated for high LLs in a smooth random potential \cite{Raikh93}.

The analytic formulae for oscillating $E_F$ that keeps the $n_e$ constant are presented in eqs.\ (\ref{deltaL}) and (\ref{deltaSin}) for the Lorentzian DOS and the sinusoidal DOS, respectively. The use of the oscillating $E_F$ has been found to be crucial for the agreement between the line shapes of calculated DOS and the experimental SdH oscillation.

\section*{Acknowledgment}
The authors would like to thank Prof. P. T. Coleridge for useful comments and for informing us of Ref. \citen{Coleridge94I}. This work was supported by Grant-in-Aid for Scientific Research (C) (18540312) and (A) (18204029) from the Ministry of Education, Culture, Sports, Science and Technology (MEXT).

\appendix
\section{Evaluation of Eq.\ (\ref{DOSsinosc}) with Eq.\ (\ref{deltaL})}
The saw-tooth variation of $E_\mathrm{F}$ results from the gradual decrease in $E_\mathrm{F}$ with decreasing $B$ (i.e., increasing $\nu$) while a (spin-degenerate pair of) LL is being filled, alternating with the sudden jump up to the next LL at (even) filling factors where the LL that has been the host of the $E_F$ is filled up. The line shape of $\Delta E_\mathrm{F}$, particularly the amplitude (the jump), is primarily determined by the Landau fan diagram eq.\ (\ref{EN}) and the width of each LL, and the detail of the LL line shape plays only minor role of slightly altering the shape of the saw teeth. This explains the resemblance between Figs.\ \ref{calcLorentz}(c) and \ref{calcsin}(c). Hence it seems to be an acceptable approximation to use $\Delta E_\mathrm{F}$ derived from another type of the DOS\@. Equation (\ref{deltaL}) has an advantage of being written analytically without approximation. In the following, we will show that eq.\ (\ref{DOSsinosc}) can readily be expanded up to an arbitrary order in $\lambda = \exp(-2 \pi \gamma)$ by employing eq.\ (\ref{deltaL}) instead of eq.\ (\ref{deltaSin}). The result lends itself to provide a simple unified view on the role of the oscillating $\Delta E_\mathrm{F}$.

First, we rewrite eq.\ (\ref{costan}) by using the relation $\tanh(\pi \gamma) = (1 - \lambda) / (1 + \lambda)$,
\begin{eqnarray}
\lefteqn{\cos (2\pi \varepsilon _F ) = } \nonumber \\
& \displaystyle{ \frac{1}{\lambda }\left[ {1 - \frac{{1 + \lambda \cos (2\pi \varepsilon_\mathrm{F}^0 )}}{{1 + \lambda ^2  + 2\lambda \cos (2\pi \varepsilon_\mathrm{F}^0 )}}} \right] + \lambda \frac{{1 + \lambda \cos (2\pi \varepsilon_\mathrm{F}^0 )}}{{1 + \lambda ^2  + 2\lambda \cos (2\pi \varepsilon_\mathrm{F}^0 )}} }. \nonumber \\
\end{eqnarray}
By employing an identity
\begin{equation}
\sum\limits_{n = 0}^\infty  {a^n \cos (n\theta ) = \frac{{1 - a\cos \theta }}{{1 + a^2  - 2a\cos \theta }}}, \nonumber
\end{equation}
we obtain
\begin{eqnarray}
\lefteqn{\cos (2\pi \varepsilon_\mathrm{F}) = \cos (2\pi \varepsilon_\mathrm{F}^0 )} \nonumber \\
&  + \sum\limits_{k = 1}^\infty  ( - 1)^k \left\{ \cos \left[2\pi (k + 1)\varepsilon_\mathrm{F}^0 \right] - \cos \left[2\pi (k - 1)\varepsilon_\mathrm{F}^0 \right] \right\} \lambda ^k \nonumber \\
\end{eqnarray}
to be replaced in eq.\ (\ref{DOSsinosc}). If we pick out the main contribution (the lowest order in $\lambda$) for each harmonic, we get 
\begin{equation}
\frac{\Delta D (\varepsilon_\mathrm{F})}{D_0} \simeq -2 \sum\limits_{k = 1}^\infty ( - 1)^k \cos \left(2\pi k \varepsilon_\mathrm{F}^0 \right)  \lambda^k
\end{equation}
or
\begin{equation}
D(E_\mathrm{F}) \simeq D_0 \left\{1 - 2 \sum\limits_{k = 1}^\infty  \cos \left(2 \pi k \frac{E_\mathrm{F}^0}{\hbar \omega_\mathrm{c}} \right) \exp \left( -2 \frac{\pi \Gamma_\mathrm{L}}{\hbar \omega_\mathrm{c}} k \right) \right\},
\end{equation}
which is nothing but a shifted and inverted Lorentzian DOS (see the discussion in the first paragraph of \S \ref{subsecLorentzian}). Therefore, to summarize, the oscillating $\Delta E_\mathrm{F}$ turns the Lorentzian DOS into a sinusoidal line shape, as described in \S \ref{subsecLorentzian} [eq.\ (\ref{DEFLor})], and, in turn, transforms the sinusoidal DOS into an inverted Lorentzian line shape.

\bibliography{twodeg,lsls,ourpps,expDOS,noteSdHHH}

\begin{thebibliography}{10}
\bibitem{HigherHarm}
After the submission of the manuscript, we became aware of Ref.\
  \citen{Coleridge94I}, which reports the amplitudes of higher harmonics. In
  the paper, however, the phases of the harmonics are not discussed.
\bibitem{NoteCmu}
The value of $C$ as well as of $\mu_\mathrm{Q}$ is found to vary from sample to
  sample, or between different conditions for the same sample. Reference
  \citen{Coleridge91} reports that the correct value is $C =$ 4 and that the
  deviation from the value is attributable to the inhomogeneity in $n_e$.
\end{thebibliography}

\end{document}